# Automated Spreadsheet Development

Angus Dunn
(angus@thedowerhouse.fsnet.co.uk)

*Few major commercial or economic decisions are made today which are not underpinned by analysis using spreadsheets. It is virtually impossible to avoid making mistakes during their drafting and some of these errors remain, unseen and uncorrected, until something turns the spotlight on them. By then it may be too late. The challenge is to find a way of creating spreadsheets which will preserve the benefit of their power and flexibility while making their creation more transparent and safer. Full documentation and documented version and quality control, section by section, of the eventual spreadsheet would be a bonus. And if the whole process could be made quicker, too, that would be a further bonus.*

## 1      WHERE WE ARE NOW

Practitioners know from bitter experience how difficult it is to produce spreadsheets that are wholly reliable. Professionals have pointed out the high proportion of spreadsheets that contain material errors [KPMG 1997]. Academic studies indicate that virtually all spreadsheets contain errors [Panko 1998, revised 2008]. [Pettifor 2003] was confident in saying that "the world is not falling apart through spreadsheet errors" - in 2003 it wasn't falling apart. In 2009 things are rather different and in due course history will allocate the proportion of blame attributable to bad spreadsheeting. The financial sector, at the root of the collapse, is a major user of spreadsheets. "Within certain large sectors [of the financial sector] spreadsheets play a role of such critical importance that without them, companies and markets would not be able to operate as they do at present" and "Spreadsheets have been shown to be fallible, yet they underpin the operation of the financial system" [Croll 2005].

Spreadsheets tend also to suffer from lack of detailed specification and poor maintainability [Pryor 2004], lack of comprehensive (or, indeed, of any) documentation, lack of modular design and inadequate testing, to name but a few.

Even if history holds bad spreadsheeting entirely blameless for the present mayhem, the fact that spreadsheeting is so widely used while at the same time so error prone is a strong indication that if a way could be found to improve reliability, global GNP would step noticeably upwards.

In a quarter century as a project banker working with and alongside spreadsheets, the author became increasingly frustrated at the failure of new tools to emerge to improve their reliability. As a result, he set out to develop one himself.

# 2        THE AIMS

A spreadsheet should be easy to read [Raffensperger 2001]. A brief glance should be enough to indicate to the reader where the inputs are, where the calculations are, where the internal checks are and which pieces contain the outputs designed to be printed out as hard copy.

The inputs and the calculations should be modular [Grossman and Özlük 2004] , each section representing a logically discrete portion of the total computation. The subject matter of each section should be clear, and it should be obvious to the reader as he scrolls through which section he is looking at at any time.

Related formulae should be in physical proximity; and the spreadsheet should read from left to right and from top to bottom [Raffensperger 2001]. In the calculations and inputs, related topics should where possible be recorded adjacent to each other; inputs should be recorded in the same order as are the calculations which draw on them.

If there is the slightest doubt about how the spreadsheet treats each section of the analysis, or if the spreadsheet uses one of a number of possible assumptions about how a particular item or set of items is calculated, this should be documented; the spreadsheet will generally be more readable if there is also an appropriate level of descriptive documentation as an aid to finding the way around.

If a data book will ultimately be required, any text entries which are required to describe individual sections should be drafted at the same time as the sections themselves.

There should be a record of who drafted each section and when.

Each section, when first drafted and after every change, should be independently checked; documentation should show whether this has been done, by whom and when.

There are no prizes for needless reinvention of the wheel; logic once written and proven to work should be available for use within the same firm in future spreadsheets which analyse the same circumstances, until that logic is itself superseded.

The spreadsheet should have a detailed specification.

The spreadsheet should be maintainable.

The hard coded parameters of the spreadsheet - start date, width, periodicity - should be input at the very end of the process, not at the very start, thus allowing these to be varied at will, without requiring changes to the template.

Finally, it goes without saying that the rules generally accepted by practitioners for safer spreadsheeting should be followed. Inputs, calculations and reports should be separated, multiple scenarios of inputs should be capable of being run through the same computation and there must be confidence that all inputs are brought – once and once only - into the calculations [Grossman and Özlük 2004], use of constants should be discouraged in formulae, all formulae in the calculations areas of the spreadsheet must be left to right consistent, there must be extensive use of self checking, circularity should be avoided where possible, and controlled and tested for convergence of result where avoidance is impossible.

## 3    THE ANSWER

The author's answer, provisionally called RingtailXL, is a Windows application which provides a framework for generating spreadsheets.  Once its template is complete, the spreadsheet, its databook and its complete specification can all be generated.

The advantages claimed for RingtailXL, in comparison with handcrafting spreadsheets directly, include:

- Substantial saving of time

- A big increase in flexibility

- A big increase in transparency

- Radical improvements in quality control

- The ability to maintain complete documentation of all the constituents of final spreadsheets

- Enforcement of good practice in the separation of data, computation and output and enforcement of left to right consistency in formulae

Together, these should result in far fewer errors.

If these claims prove robust, major spreadsheet users would experience radical improvements in productivity, consistency and quality.

## 4    HOW

RingtailXL enables users to assemble templates to generate spreadsheet models.  The templates consist of three different elements, called Components, Skeletons and Models.

A Component is a representation of a small group of adjacent rows to be placed on one of the calculation sheets of the final spreadsheet, designed to model a particular aspect.  Indeed, when the final spreadsheet is generated, these lines are shown "Grouped".

**From this . . .**

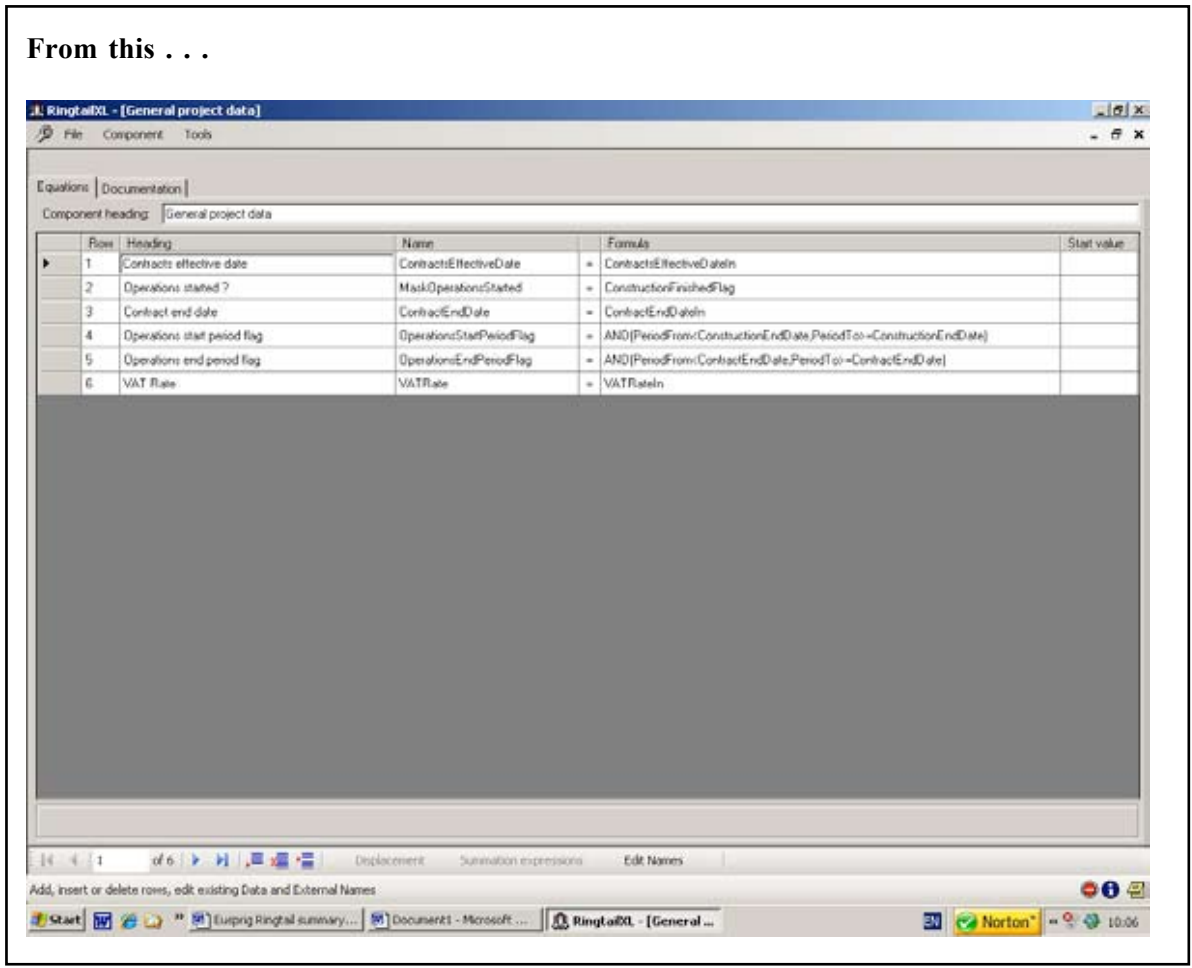

A Component may contain one or more "Embedded Components", thereby allowing the same logic to be used several times but with different inputs or a complex task to be broken into a number of sub-tasks.

Once sufficient Components have been prepared to undertake all the tasks the final spreadsheet needs, these are strung together into a single "Skeleton" of the calculation area(s) of the final spreadsheet;  it is the collection of these separate Components which provide the resulting spreadsheet with its "modularity".

Additionally, in the Skeleton, the structure of each data input, the width of all the formulae (ie, full width, single column or special) and the checks which the spreadsheet will undertake, are all defined.  If many models of similar structure are made for each particular type of project, a unique skeleton would probably be constructed for each such type.

The Model imports a single Skeleton, and requires the user to complete the remaining structural details - details of the data structures, all special formula widths, and display formats for the output of each formula - and enables the user to enter multiple data scenarios and to define output sheets, in the form of reports and charts.

It also enables the user to define additional calculation sheets, and specify on which such sheet each formula in the spreadsheet will appear  Thus, though the default structure of the resulting Workbook is for all the calculations to be placed on a single worksheet called "Workings", additional calculation sheets can be generated by users who prefer a different style.  The system enforces  separation into one or more input sheets, one or more calculation sheets, a sheet of checks and sheets which report output, in the form both of reports and charts.

 

**. . . via this . . .**

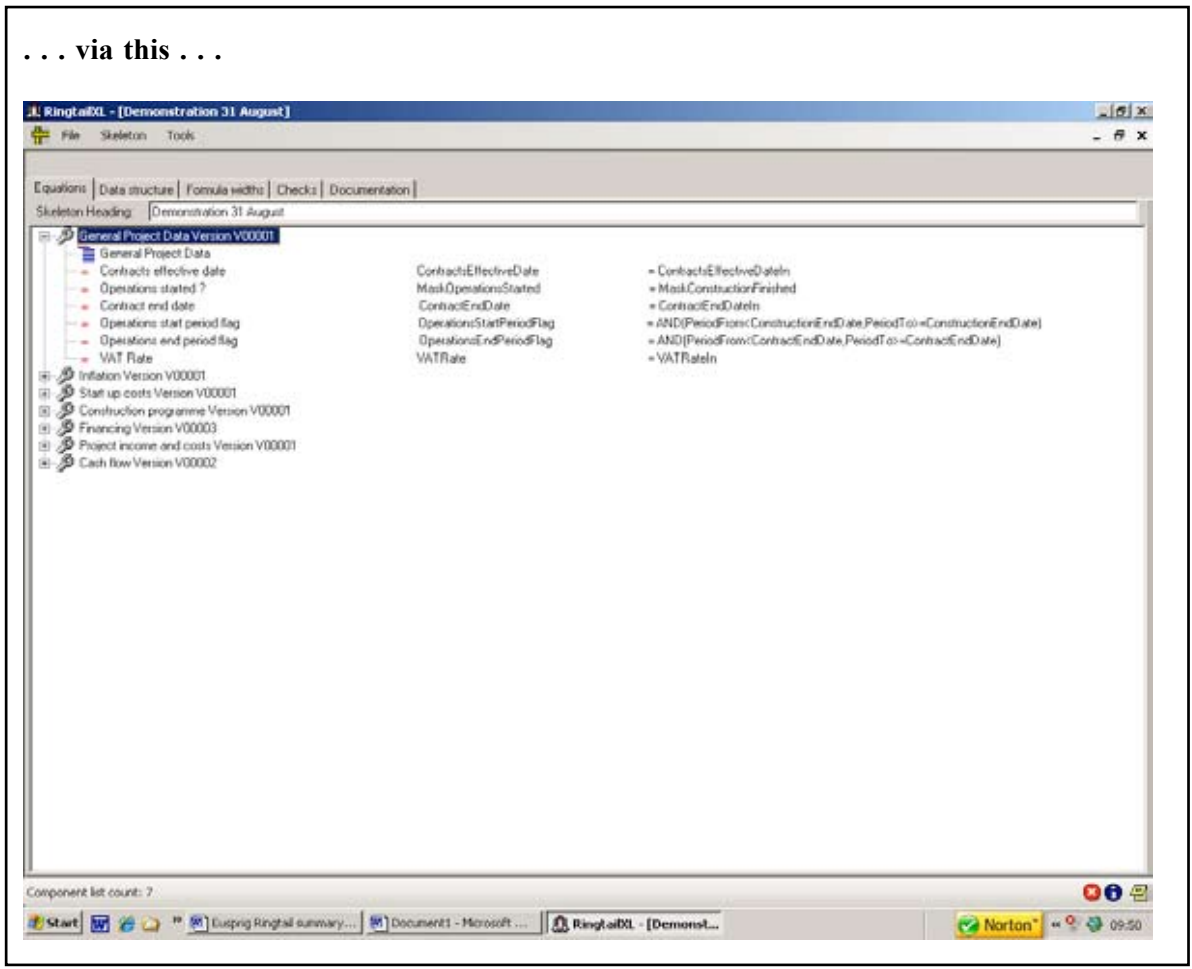

From a higher level, an element at a lower level can be drilled into, changed, saved as a new version and the new version substituted for the old version in the element at the level above. Thus from a Model, its Skeleton can be called up and changed, and from a Skeleton any Component can be called up and changed, and from a Component, any Embedded Component can be called up and changed.  After returning to the level above, information and parameters stored at that level which are still relevant are retained;  information which is no longer relevant is discarded.

Each element has a feature allowing it to be diagnosed for completeness and for certain errors.

From the Model, a full Excel Workbook, a detailed specification and a databook can all be generated.

The whole spreadsheet development cycle should be undertaken within RingtailXL; during its lifetime many versions of the spreadsheet will probably be generated, most of which will be discarded.  Provided the templates have been kept, any version of the spreadsheet can be re-created at any time, together with its specification and databook.  The intention is that manual changes are never made to the  spreadsheets themselves and indeed these all contain a feature which shows whether each is "clean" or has been changed by hand since it was generated.

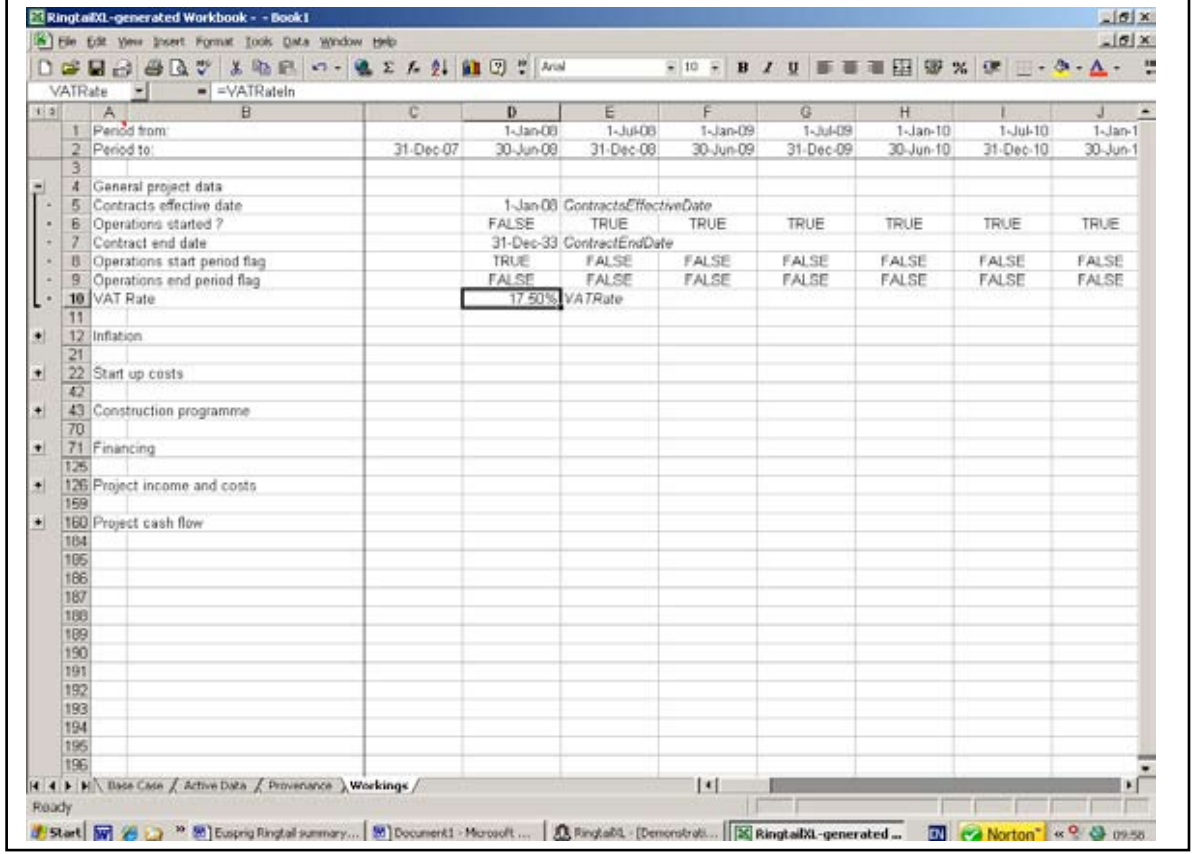

## 5 DOCUMENTATION

Each element (the Components, the Skeletons and the Models) stores four pieces of documentation.

"Notes" and "Data Book Entry" are input by the user. The Notes can be used for any purpose. The Data Book Entry shows what a Data Book, generated at the end, will contain in respect of the element concerned.

The other two pieces of documentation, "Status" and "Audit Trail" are generated by the system.

The Status has three possible values – OK (this element has not been materially changed since it was last checked), Warning (this element has been materially changed on one occasion since it was last checked) and Failure (this element has been changed from one which previously showed a Warning or a Failure). The Status of any element also records the Status of all elements contained within it. Icons on the main RingtailXL window indicate the Status of the currently active element.

The Audit Trail shows the version from which the element was derived and the version as which it has been saved and lists all the changes (whether considered material by the system or not) between the two versions.

## 6    CONCLUSIONS

The work on RingtailXL leads to the conclusion that it will be possible to use spreadsheet generators to draft spreadsheets that are modular, easy to read, fully documented, fully specified, demonstrably checked, and which need no maintenance.  The work also holds out the hope that the process of developing such spreadsheets will be less stressful and faster than is the process of hand-crafting directly into Excel.

Whether the ideas contained within RingtailXL will take off as the basis of such a spreadsheet generator remains to be seen, but they certainly could.  RingtailXL is actively looking for interested partners to help take it into the mainstream.

When such Workbook generators are generally used, they will make the world a safer place. And not only will they make it safer – by increasing the reliability of the tool almost universally used to underpin decision making, they will make it richer too.  Which will perhaps offset to an extent the blame history is likely to lay at spreadsheeting's door for the way the world is in 2009.